\documentstyle[twoside,psfig]{article}

\catcode`\@=11
\long\def\@makefntext#1{
\protect\noindent \hbox to 3.2pt {\hskip-.9pt
$^{{\eightrm\@thefnmark}}$\hfil}#1\hfill}       

\def\@makefnmark{\hbox to 0pt{$^{\@thefnmark}$\hss}}    

\def\ps@myheadings{\let\@mkboth\@gobbletwo
\def\@oddhead{\hbox{}
\rightmark\hfil\eightrm\thepage}
\def\@oddfoot{}\def\@evenhead{\eightrm\thepage\hfil
\leftmark\hbox{}}\def\@evenfoot{}
\def\sectionmark##1{}\def\subsectionmark##1{}}

\oddsidemargin=\evensidemargin
\newcounter{sectionc}\newcounter{subsectionc}\newcounter{subsubsectionc}
\renewcommand{\section}[1] {\vspace{12pt}\addtocounter{sectionc}{1}
\setcounter{subsectionc}{0}\setcounter{subsubsectionc}{0}\noindent
    {\tenbf\thesectionc. #1}\par\vspace{5pt}}
\renewcommand{\subsection}[1] {\vspace{12pt}\addtocounter{subsectionc}{1}
    \setcounter{subsubsectionc}{0}\noindent
    {\bf\thesectionc.\thesubsectionc. {\kern1pt \bfit #1}}\par\vspace{5pt}}
\renewcommand{\subsubsection}[1] {\vspace{12pt}\addtocounter{subsubsectionc}{1}
    \noindent{\tenrm\thesectionc.\thesubsectionc.\thesubsubsectionc.
    {\kern1pt \tenit #1}}\par\vspace{5pt}}
\newcommand{\nonumsection}[1] {\vspace{12pt}\noindent{\tenbf #1}
    \par\vspace{5pt}}

\newcounter{appendixc}
\newcounter{subappendixc}[appendixc]
\newcounter{subsubappendixc}[subappendixc]
\renewcommand{\thesubappendixc}{\Alph{appendixc}.\arabic{subappendixc}}
\renewcommand{\thesubsubappendixc}
    {\Alph{appendixc}.\arabic{subappendixc}.\arabic{subsubappendixc}}

\renewcommand{\appendix}[1] {\vspace{12pt}
        \refstepcounter{appendixc}
        \setcounter{figure}{0}
        \setcounter{table}{0}
        \setcounter{lemma}{0}
        \setcounter{theorem}{0}
        \setcounter{corollary}{0}
        \setcounter{definition}{0}
        \setcounter{equation}{0}
        \renewcommand{\thefigure}{\Alph{appendixc}.\arabic{figure}}
        \renewcommand{\thetable}{\Alph{appendixc}.\arabic{table}}
        \renewcommand{\theappendixc}{\Alph{appendixc}}
        \renewcommand{\thelemma}{\Alph{appendixc}.\arabic{lemma}}
        \renewcommand{\thetheorem}{\Alph{appendixc}.\arabic{theorem}}
        \renewcommand{\thedefinition}{\Alph{appendixc}.\arabic{definition}}
        \renewcommand{\thecorollary}{\Alph{appendixc}.\arabic{corollary}}
        \renewcommand{\theequation}{\Alph{appendixc}.\arabic{equation}}
        \noindent{\tenbf Appendix \theappendixc #1}\par\vspace{5pt}}
\newcommand{\subappendix}[1] {\vspace{12pt}
        \refstepcounter{subappendixc}
        \noindent{\bf Appendix \thesubappendixc. {\kern1pt \bfit #1}}
    \par\vspace{5pt}}
\newcommand{\subsubappendix}[1] {\vspace{12pt}
        \refstepcounter{subsubappendixc}
        \noindent{\rm Appendix \thesubsubappendixc. {\kern1pt \tenit #1}}
    \par\vspace{5pt}}

\newcommand{\textlineskip}{\baselineskip=13pt}
\newcommand{\smalllineskip}{\baselineskip=10pt}

\def\eightcirc{
\begin{picture}(0,0)
\put(4.4,1.8){\circle{6.5}}
\end{picture}}
\def\eightcopyright{\eightcirc\kern2.7pt\hbox{\eightrm c}}

\newcommand{\copyrightheading}[1]
    {\vspace*{-2.5cm}\smalllineskip{\flushleft
    {\footnotesize International Journal of Modern Physics D #1}\\
    {\footnotesize $\eightcopyright$\, World Scientific Publishing
     Company}\\
     }}

\newcommand{\publisher}[2]{{\begin{center}\footnotesize\smalllineskip
    Received #1\\
    Revised #2
    \end{center}
    }}

\def\abstracts#1#2#3{{
    \centering{\begin{minipage}{4.5in}\footnotesize\baselineskip=10pt
    \parindent=0pt #1\par
    \parindent=15pt #2\par
    \parindent=15pt #3
    \end{minipage}}\par}}

\renewenvironment{thebibliography}[1]
    {\frenchspacing
     \ninerm\baselineskip=11pt
     \begin{list}{\arabic{enumi}.}
    {\usecounter{enumi}\setlength{\parsep}{0pt}
     \setlength{\leftmargin 12.7pt}{\rightmargin 0pt} 
     \setlength{\itemsep}{0pt} \settowidth
    {\labelwidth}{#1.}\sloppy}}{\end{list}}

\newcounter{itemlistc}
\newcounter{romanlistc}
\newcounter{alphlistc}
\newcounter{arabiclistc}

\newcommand{\fcaption}[1]{
        \refstepcounter{figure}
        \setbox\@tempboxa = \hbox{\footnotesize Fig.~\thefigure. #1}
        \ifdim \wd\@tempboxa > 5in
           {\begin{center}
        \parbox{5in}{\footnotesize\smalllineskip Fig.~\thefigure. #1}
            \end{center}}
        \else
             {\begin{center}
             {\footnotesize Fig.~\thefigure. #1}
              \end{center}}
        \fi}

\newcommand{\tcaption}[1]{
        \refstepcounter{table}
        \setbox\@tempboxa = \hbox{\footnotesize Table~\thetable. #1}
        \ifdim \wd\@tempboxa > 5in
           {\begin{center}
        \parbox{5in}{\footnotesize\smalllineskip Table~\thetable. #1}
            \end{center}}
        \else
             {\begin{center}
             {\footnotesize Table~\thetable. #1}
              \end{center}}
        \fi}

\def\@citex[#1]#2{\if@filesw\immediate\write\@auxout
    {\string\citation{#2}}\fi
\def\@citea{}\@cite{\@for\@citeb:=#2\do
    {\@citea\def\@citea{,}\@ifundefined
    {b@\@citeb}{{\bf ?}\@warning
    {Citation `\@citeb' on page \thepage \space undefined}}
    {\csname b@\@citeb\endcsname}}}{#1}}

\newif\if@cghi
\def\cite{\@cghitrue\@ifnextchar [{\@tempswatrue
    \@citex}{\@tempswafalse\@citex[]}}
\def\citelow{\@cghifalse\@ifnextchar [{\@tempswatrue
    \@citex}{\@tempswafalse\@citex[]}}
\def\@cite#1#2{{$\null^{#1}$\if@tempswa\typeout
    {IJCGA warning: optional citation argument
    ignored: `#2'} \fi}}

\def\pmb#1{\setbox0=\hbox{#1}
    \kern-.025em\copy0\kern-\wd0
    \kern.05em\copy0\kern-\wd0
    \kern-.025em\raise.0433em\box0}

\def\fnt#1#2{\footnotetext{\kern-.3em
    {$^{\mbox{\scriptsize #1}}$}{#2}}}

\def\@makefnmark{\hbox to 0pt{$^{\@thefnmark}$\hss}}    

\def\ps@myheadings{%
    \let\@oddfoot\@empty\let\@evenfoot\@empty
    \def\@evenhead{\slshape\leftmark\hfil}
    \def\@oddhead{\hfil{\slshape\rightmark}}
    \let\@mkboth\@gobbletwo
    \let\sectionmark\@gobble
    \let\subsectionmark\@gobble
    }
\font\tenrm=cmr10
\font\tenit=cmti10
\font\tenbf=cmbx10
\font\bfit=cmbxti10 at 10pt
\font\ninerm=cmr9

\font\eightrm=cmr8

\def\qed{\hbox{${\vcenter{\vbox{            
   \hrule height 0.4pt\hbox{\vrule width 0.4pt height 6pt
   \kern5pt\vrule width 0.4pt}\hrule height 0.4pt}}}$}}

\begin{document}
\setlength{\textheight}{7.7truein}  

\thispagestyle{empty}

\markboth{\protect{\footnotesize\it Quark-diquark equations of
state models  $\ldots$}}{\protect{\footnotesize\it  Quark-diquark equations of
state models $\ldots$}}

\normalsize\textlineskip

\setcounter{page}{1}

\copyrightheading{} 

\vspace*{0.88truein}

\centerline{\bf QUARK-DIQUARK EQUATIONS OF STATE MODELS:}
\vspace*{0.035truein}
\centerline{\bf THE ROLE OF INTERACTIONS}
\vspace*{0.37truein}

\centerline{\footnotesize   J. E. HORVATH\footnote{Email: foton@astro.iag.usp.br} ,  G. LUGONES\footnote{Email: glugones@astro.iag.usp.br} }
\baselineskip=12pt
\centerline{\footnotesize\it Instituto Astron\^omico e Geof\'{\i}sico, Universidade de S\~ao Paulo,}
\baselineskip=10pt
\centerline{\footnotesize\it Rua do Mat\~ao 1226, Cidade Universit\'aria 05508-900 S\~ao Paulo SP, Brazil}
\vspace*{10pt}
\vspace*{0.225truein}
\centerline{ and }
\vspace*{10pt}
\centerline{\footnotesize   J.A. DE FREITAS PACHECO \footnote{Email: pacheco@obs-nice.fr}}
\baselineskip=12pt
\centerline{\footnotesize\it Observatoire de la C\^ote d'Azur,}
\baselineskip=10pt
\centerline{\footnotesize\it Departement Augustin Fresnel, BP 4229, 06304 Nice Cedex 4, France}
\vspace*{10pt}
\vspace*{0.225truein}
\publisher{(received date)}{(revised date)}

\vspace*{0.21truein}
\abstracts{Recent observational data suggests a high compacticity (the
quotient $M/R$) of some
"neutron" stars. Motivated by these works we revisit models based on
quark-diquark degrees of freedom and address the question of whether
that matter is stable against diquark disassembling and
hadronization within the different models.
We find that equations of state modeled as effective
$\lambda \phi^{4}$ theories do not generally produce stable self-bound
matter and are not suitable for constructing very compact star models,
that is the matter would decay into neutron matter.
We also discuss some insights obtained by including
hard sphere terms in the equation of state to model repulsive
interactions.
We finally compare the resulting equations of state with previous models
and emphasize the role of the boundary conditions at the surface of
compact self-bound stars, features of a possible normal crust of the latter
and related topics.}{}{}

\vspace*{1pt}\textlineskip  
\section{Introduction}  
\vspace*{-0.5pt}
\noindent
Recent works on compact stars have attracted the attention
because not only the masses may be determined, but also
indications of the radii are available for the first time. In some cases
the compactness of the sources has
been claimed, athough these results have yet to be confirmed carefully.
Nonetheless, it is worthwile to entertain the possibility that at least
some compact stars are extremely compact, or in other words, that their
radii are $\sim 30-40 \%$ less than the "canonical" $10 \, km$ favored by
neutron matter models for $M \, \sim 1 \, M_{\odot}$.

Claims of high compactness
from the analysis of the binary Her X-1 \cite{li,Dey98}($M = 0.98 \pm 0.12 M_{\odot}$ and $R =
6.7 \pm 1.2$ km) and  of the isolated nearby RX
J185635-3754 \cite{Pons} ($M \approx 0.9 M_{\odot}$ and $R \approx 6$ km ) have been made. In both cases,
the results have been revisited and challenged by other groups \cite{Rey,Kaplan} which found
radii in the ballpark of conventional neutron star models. This stresses the cautionary remarks
made by several researchers about the high-compactness objects and guarantees further
studies, already undertaken in most cases.
We should add that Li et al. \cite{Li99b} have also added an argument
about the source 4U 1728-34, showing that
conventional accretion models would indicate a very compact source, as
quantified by their limit in the mass-radius plane. Again the actual distance to the source is a
matter of concern.
If really present in these sources compactness would be
extremely difficult (and perhaps impossible) to model using underlying
equations of state based on ordinary hadrons alone, and a natural alternative would
be to consider deconfined matter (or other component, like kaon matter or hyperons).
Other stars that have been claimed to be made up of deconfined matter
are the compact objects associated with the X-ray bursters GRO J1744-28
\cite{Cheng98}, and SAX J1808.4-3658 \cite{Li99a}.

It is widely believed that at high temperatures and/or
high baryon number densities deconfinement of color is achieved. It is not
very clear how high the densities/temperatures must be because of the
extensively discussed failure of perturbative expansions around the
transition point. The physical realization of deconfined matter immediately
above the deconfinement point is also doubtful. The phase structure
seems to be very rich and lattice simulations
suggest that "asymptotia" is not sharply reached.
There is an expectation that diquarks (e. g. a spin-0, color-antitriplet
bound state of two quarks) might occur as a component in the QCD
plasma in these conditions (see Ref. 9 and references therein for
a review). Such diquarks would be expected to be favored by Bose statistics.
If bound, the diquark binding energy would also contribute in
making this phase favorable to ordinary uncorrelated quark matter.
At very high densities diquarks are expected to lose their identity
and must eventually dissolve into quarks, even if (as stated) there is
no consensus about the onset of the asymptotic regime.

As a working hypothesis, we shall treat diquarks as single entities,
and explore the possibility of a quark-diquark phase being
absolutely stable, quite analogously to the well known
strange quark matter hypothesis. The difference here is that we shall not invoke
a new quantum number to bind the mixture, this role is partially played by
the expected bosonic character of diquarks.

We stress that absolute stability of the mixture
does {\it not} require the diquark itself to be bound (see
Ref. 10, where a value of $m_{D} \sim 700 \, MeV$ has been claimed), but
rather that the energy per baryon of the quark-diquark mixture to be
less than the nucleon mass. The functional form of the free energy may or may not
allow this to happen, and therefore the model-dependence of the description
is the actual motive of concern, rather than the exact value of the diquark
mass at a given density. As we shall see, this is crucial within
$\lambda \phi^{4}$ theories. We shall return to this point later.
We devote the next two sections to discuss the effective models of the
quark-diquark phase and present a critical appraisal in the last section.

\section{Bag-inspired quark-diquark Model}

\subsection{The equation of state}

Some time ago it was the interesting controversy about the compactness of
Her X-1 that prompted two of us \cite{HP} to consider models in which
diquarks are considered fundamental degrees of freedom and treated as
effective bosons. The results were derived using an effective $\lambda \phi^{4}$
model and the equation of state given in \cite{CSW}, shown to be
valid in this case. This approach goes back to Refs.\cite{Donoghue,Kastor}
in which similar bosonic diquark dynamics were employed.

In constructing the EOS for quark-diquark matter there are some quantities
that had to be treated parametrically: the diquark mass $m_{D}$, a vacuum
energy density $B$
and the coupling constant $\lambda$ . In fact, the effective character of
this description
allows for other possibilities than the previously selected
($m_{D} = 575 MeV$, $B = 57 MeV fm^{-3}$, $\lambda = 27.8$) to be
considered.

Following the approach of Ref. 11
we have modeled the quark - diquark plasma at zero temperature as a mixture
of a free Fermi
gas of $u$ and $d$ quarks, and a $ud$ diquark gas treated in the framework
of a $\lambda \phi^4$
theory. Confinement is introduced by hand by means of a phenomenological bag
constant $B$.

As shown by Donoghue et al. \cite{Donoghue} a $\lambda \phi^4$ theory
leads to a quadratic expression for the pressure of a pure diquark gas in
the low-density regime
(valid for densities up to $\sim 10 \times \rho_0$, being $\rho_0$ the
nuclear saturation density \cite{Kastor}):

\begin{equation}
P_D = \frac{\lambda}{2 m_D^2} n_D^2
\end{equation}

\noindent whereas the energy density is given simply by a rest mass term
$m_D n_D$.
Therefore, the full EOS is quite simple and given by

\begin{equation}
P = \frac{\lambda}{2 m_D^2} n_D^2 + \frac{1}{5} \frac{\pi^{4/3} \hbar }{m_q}
( n_{u}^{5/3} + n_{d}^{5/3} ) - B,
\end{equation}

\begin{equation}
\epsilon = m_u n_u + m_d n_d + m_D n_D + \frac{3}{10}
\frac{\pi^{4/3} \hbar }{m_q } ( n_{u}^{5/3} + n_{d}^{5/3} ) + B,
\end{equation}

\noindent
and must be subject to the following conditions:

1) chemical equilibrium between diquarks and quarks,
$D \rightleftharpoons u + d$, which requires $\mu_D = \mu_u + \mu_d$

\begin{equation}
\frac{\epsilon_D + P_D}{n_D} = \mu_u + \mu_d ,
\end{equation}

2) bulk electrical charge neutrality $ (1/3) n_D + (2/3) n_u - (1/3) n_d = 0$,

3) baryon number conservation $n_B = (2/3) n_D + (1/3) (n_u + n_d)$.

Once the parameters $B$, $\lambda$ and $m_D$ are given, the equation of
state can be calculated for a given value of, for example, $n_D$.
Note that we did not include strange quarks in the admixture, thus
ignoring  CFL phases \cite{Al}. Rather than extrapolating down in density
(as done for CFL models), we work "bottom up" here
by postulating a gradual lose of correlations (diquarks)
starting from the deconfinement point.

\subsection{Effective hard-sphere model : interactions in the quark-diquark
mixture}

The effective models of a quark-diquark mixture based on a $\lambda
\phi^{4}$ theory
do not say much about some fundamental questions. One of
these is the strength of the repulsion between particles, which is closely
related to their size and must be guessed by resorting to experimental data.
Alternatively,
the interactions may be modeled with a potential at relatively low
densities, the
latter being much more amenable to well-known analysis.
It would be desirable to have a better insight to this and related questions
since, in principle, the diquark size and scattering behavior are
measurable (see Ref. 9 and references therein). We can account for
this by introducing a
scattering length in both the pressure and energy density of the Bose gas
\cite{Lee,Rapp}.
Thus, we can consider the following expresions for the pressure and
the energy density of diquarks

\begin{equation}
P_D = \frac{4 \pi a}{m_D} n_D^2
\end{equation}

\begin{equation}
\epsilon_D = \frac{4 \pi a}{m_D} n_D^2 \bigg(1 + \frac{128}{15 \pi^{1/2}}
(a^3 n_D)^{1/2} \bigg)
\end{equation}

As shown below quantitatively (and expected qualitatively),
the repulsive interactions makes the quark-diquark mixture less favourable
energetically than the "optimal" Fermi-Bose mixture (see also Ref. 17).

As the diquark pressure depends on $n_D^2$ in both the $\lambda \phi^4$-based EOS
and the approximation of the present section, we can establish immediately
a relation between the
coupling constant $\lambda$ and the scattering length $a$ by equating the pressure
in both approaches. This gives

\begin{equation}
\lambda = 30.6 \times \bigg( \frac{a}{0.4 fm} \bigg)  \bigg( \frac{m_D}{600 MeV} \bigg).
\end{equation}

It is interesting to note that although the comparison is very crude, it gives a kind of
self-consistent result:
the scattering length reflects the diquark size, which is expected to be of the order of
the typical instanton radius, i.e. $1/3$ fm \cite{Rapp}.  For a "reasonable" diquark mass
of $600 \, MeV$ the above formula gives a value of the
coupling constant which is very close to the
claimed value \cite{Donoghue} of $\lambda= 27.8$ .

\section{Stability of quark-diquark matter}

Different compositions for diquark matter have been analyzed in the
literature. Although it is not clear at all whether any of these compositions can
actually happen in neutron stars, a mixture of quarks $u$ and $d$ and $ud$ diquarks
is perhaps the simplest possibility since it
may be thought to arise from pure neutron matter deconfinement, and
therefore it seems a good starting point for our analysis. Weak interaction decays
must alter this simplest composition and remains to be studied in detail.

It is usually assumed as a criterion for stability of a given deconfined EOS
that the energy per baryon of the deconfined phase
(at $P=0$ and $T=0$) must be lower than $939 \, MeV$ (the neutron mass).
The effect of considering a scattering length amplitude $a$
(or, equivalently, a positive value of $\lambda$) makes
more difficult for the quark-diquark matter to fulfill this condition and therefore
to be absolutely stable.

With the above condition we can determine the set of parameters
$B$, $m_D$ and $\lambda$  (or $a$)
that gives a stable EOS (see Figures 1 and 2).
Except for a marginal choice of the parameters (e. g. $B \sim
10 \, MeV \, fm^{-3}$, $\lambda \sim 10$ and very
small diquark masses $m_D \sim 400 MeV$, the EOS
posses an energy per baryon higher than the neutron mass.
That is, it seems difficult that in the frame of a $\lambda \phi^4$ theory
quark-diquark matter could be
the ground state of strong interactions.

Another condition that has to be considered is the empirically known
stability of normal nuclear matter
against deconfinement at zero pressure. In other words the energy per baryon
of deconfined matter (a pure gas of quarks $u$ and $d$) at zero pressure and
temperature must be higher than the neutron mass. In the framework
of a MIT-based EOS it has been shown that this condition imposes that the
MIT bag constant must be greater than 57 $MeV \, fm^{-3}$   (see Ref. 18).
This is marked in Figs.1 and 2 with vertical lines.

\section{Discussion}

We have found in this work that for a wide range
of the bag constant $B$, the diquark mass $m_D$
and the coupling constant $\lambda$, the quark-diquark mixture has an
energy per baryon higher than the neutron mass.
Absolute stability is obtained only for very low values of $B$, and $m_D$, if at all.

Within this description diquarks are favoured over a
pure mixture of quarks $u$ and $d$ due to their
Bose character and eventually its hypotetical binding energy. However, we
must note, as stressed in Ref. 19, that the effect of Bose
condensation seems to be much more important for quark-diquark matter
self-binding than the binding energy of the diquark itself. In other words,
even an unbound diquark may have left room for stable quark-diquark matter,
although this is finally not the case, at least within this model.
Recently, detailed calculations of the
equation of state have been performed in a model in which confinement and
masses are related, as the quark mass-density-dependent model, finding a wide set
of parameters for which quark-diquark matter is stable \cite{vigilantes}.
Therefore the absolute stabilty of quark-diquark matter seems to be quite
model dependent.

A corollary of these findings has a big potential impact on compact star
models made up
of quark-diquark matter. Without repeating the analysis here, we just point
out that (as expected) in the frame of the $\lambda \phi^4$ model
a self-bound quark-diquark mixture indeed produces
very compact models. The mass-radius relation of quark-diquark models
 presents the same qualitative shape as the
curves found for strange stars . This is a direct
consequence of the existence of a zero pressure point at finite density
(of the order of the nuclear saturation density).
However, although they give a very compact star structure,
models based in the $\lambda \phi^4$ model are not found to be stable
against hadronization, although those based on quark mass-density dependent are
\cite{vigilantes}.
Therefore we do {\it not} claim a statement about the binding energy of
quark-diquark matter but rather a feature of the particular
model description of the equation of state.

The work of Ref. 20 has discussed how to check the
viability of a quark-diquark picture of the
source Her X-1 (claimed to be a strange star in Refs. 1 and 2
by using an approach pioneered by Vaidya and Tikekar \cite{VT}, in which
a fluid is found to support a given geometry. The proposed quark-diquark
mixture is one such a fluid and the authors calculations seem to agree with
previous works. This is
quite expected, although the proof may have helped to enlighten the
issue. The authors have criticized the boundary conditions used in former
works and also the (claimed) inconsistency between the models
of  Ref. 11 and a
deconfinement density higher than the outer densities found. However,
we believe that this is a misleading argument since the
feature happens in {\it all} self-bound
stars made out of (hypothetic) self-bound states like SQM. In the latter,
the pressure has always a zero for a non-zero value of the energy density,
and this energy density is always lower than the deconfinement density.
Therefore the boundary condition at the surface is adequate because the
radius is found by imposing $P=0$ there, and
the total energy density is non-zero in a perfectly consistent way. These
"pure" quark or quark-diquark stars without any crust are certainly more
compact than any counterpart having a normal crust. It is also well known
that there can not be in these models a coexistence of both types of matter
because the lower (exotic) energy state would swallow the normal matter
liberating energy. In quark-diquark models a possible normal crust must be
supported by electrostatic forces as well, and therefore it is the charge
density at the surface which determines the mass of the former. No electrostatically
supported crust is possible without electrons, not present here
because of the assumptions of the model. Should a more complete model been
developed (including strange quarks and detailed beta equilibrium), a tiny
crust may have been added, analogously to strange star models and with a similar
thickness $\sim \, 100 \, m$.
Given that the uncertainties in the
microscopic description of the quark-diquark phase are likely to have a much
larger effect on the radius than the presence or absence of a normal crust,
we think that this is not really an issue unless very detailed models have
to be constructed. However, the message here is that there is no inconsistency
whatsoever in the surface boundary conditions and, of course, also no
"large" crusts unless the electrostatic forces happen to be orders of magnitude
more important than for strange stars. Given the state-of-the-art of
quark-diquark models, more studies are guaranteed to see whether they may
play a role in compact star structure.

\nonumsection{Acknowledgements}
\noindent G. Lugones acknowledges
the Instituto de Astronomia Geof\'\i sica e Ci\^encias Atmosf\'ericas da
Universidade de S\~ao Paulo for its hospitality and the financial
support received from the Funda\c c\~ao de Amparo \`a Pesquisa do Estado de
S\~ao Paulo. J.E. Horvath wishes to acknowledge
the CNPq Agency (Brazil) for partial financial support.

\nonumsection{References}

\eject


\begin{figure}
\fcaption{ The energy per baryon for quark-diquark matter within
the $\lambda \phi^4$-based EOS as a function of the bag constant
$B$, for different values of the coupling constant $\lambda$. The
mass of quarks $u$ and $d$ is set to $360 \, MeV$ and the diquark
mass to $m_D \, = \, 400 MeV$. In order to be absolutely stable
the energy per baryon must be lower than $939 \, MeV$ (horizontal
line) and the bag constant must be greater than $57 \, MeV \,
fm^{-3}$ (vertical line) . This condition is never fulfilled. If
we relax the condition imposed on $B$ there is a small region in
the parameter space where stability is allowed.}
\end{figure}

\begin{figure}
\fcaption{ The same as the previous figure but for $m_D \, = \,
600 \, MeV$. Absolute stability is never fulfilled even if we
relax the condition imposed on $B$. }
\end{figure}


\noindent
\begin{thebibliography}{99}

\bibitem{li} X. -D. Li, Z. -G. Lai and Z. -R. Wang, {\it Astron. Astrophys.}
303, (1995) L1

\bibitem{Dey98} M. Dey, I. Bombaci, J. Dey, S. Ray and B.C. Samanta,
{\it Phys. Lett. } B438, (1998) 123; Addendum (1999), B447, 352; Erratum
B467, (1999) 303.

\bibitem{Pons} J. Pons, F. M. Walter, J. M. Lattimer, M. Prakash, R.
Neuh\"auser, and Penghui An., {\it Astrophys. J.}, 564, (2002) 981.

\bibitem{Rey} A.P. Reynolds; H. Quaintrell; M.D. Still; P. Roche; D. Chakrabarty and S.E. Levine,
{\it MNRAS}, 288 (1997), 43.

\bibitem{Kaplan} D. L. Kaplan, M. H. van Kerkwijk, J. Anderson, astro-ph/0111174 (2001).

\bibitem{Li99b} X. Li, S. Ray, J. Dey, M. Dey \& I. Bombaci, {\it Astrophys.
J.} 527, (1999) L51.

\bibitem{Cheng98} K. S. Cheng, Z. G. Dai, D. M. Wai and T. Lu, Science
280, 407 (1998).

\bibitem{Li99a} X. Li, I. Bombaci, M. Dey, J. Dey \& E. P. J. van den Heuvel,
{\it Phys. Rev. Lett.} 83, (1999) 3776.

\bibitem{anselmino} M. Anselmino, E. Pedrazzi, S. Ekelin, S. Fredriksson and
D.B. Litchenberg, {\it Rev. Mod. Phys.}65 (1993) 1199.

\bibitem{Hess} M. Hess, F. Karsch, E.Learmann and I. Wetzorke, {\it
Phys.Rev.} D58 (1998) 111502

\bibitem{HP} J. E. Horvath and J. A. de Freitas Pacheco,
{\it Int. J. of Mod. Phys.} D 7, (1998) 19.

\bibitem{CSW} M. Colpi, S.L. Shapiro and I. Wasserman, {\it Phys. Rev. Lett.} 57 (1986) 2485.

\bibitem{Donoghue} J. F. Donoghue and K. S. Sateesh, {\it
Phys.Rev.} D38 (1988) 360.

\bibitem{Kastor} D. Kastor and J. Traschen, {\it Phys. Rev.} D 44, (1991)
3791

\bibitem{Al} see M. Alford, invited talk to the {\it Proceedings of the Workshop
on Compact Stars and the QCD Phase Diagram}, Copenhagen 2001,
astro-ph/0110150 for a recent review.

\bibitem{Lee} T. D. Lee and C. N. Yang, Phys. Rev. 105, 1119 (1957).

\bibitem{Rapp} R. Rapp, T. Sch\"afer, E. V. Shuryak and M. Velkovsky,
Phys. Rev. Lett. 81, 53 (1998).

\bibitem{fahrijaffe} E. Farhi and R. L. Jaffe, Phys. Rev. D 30, 2379 (1984).

\bibitem{vigilantes} G. Lugones and J. E. Horvath, submitted to IJMPD.

\bibitem{hijosdemil} R. Sharma and S. Mukherjee, {\it Mod. Phys. Lett.} A 16 (2001) 1049.

\bibitem{VT} P.C. Vaidya and R. Tikekar, {\it Jour. Astron.Astrophys.}, 3
(1982), 325

\end{thebibliography}
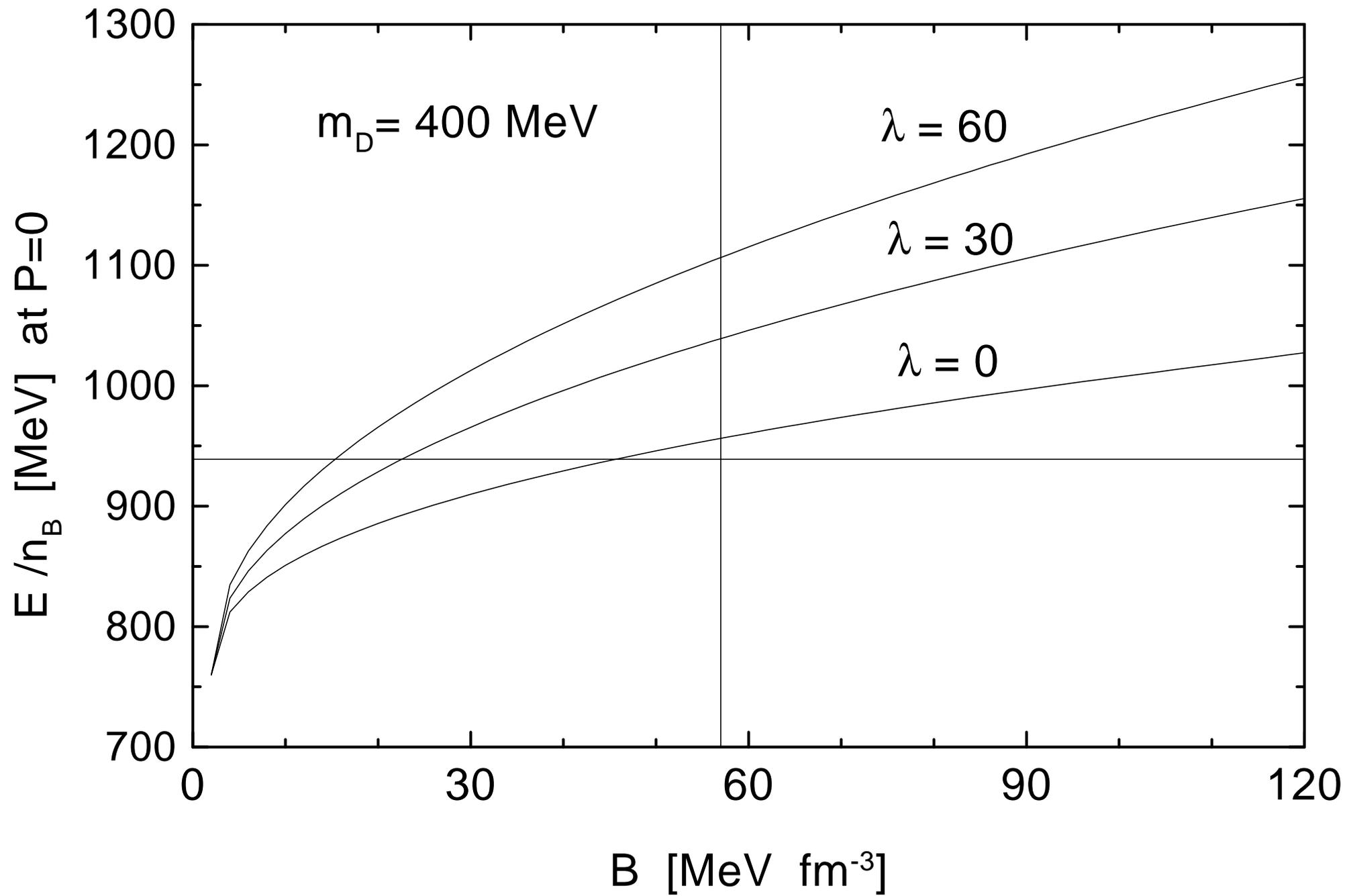
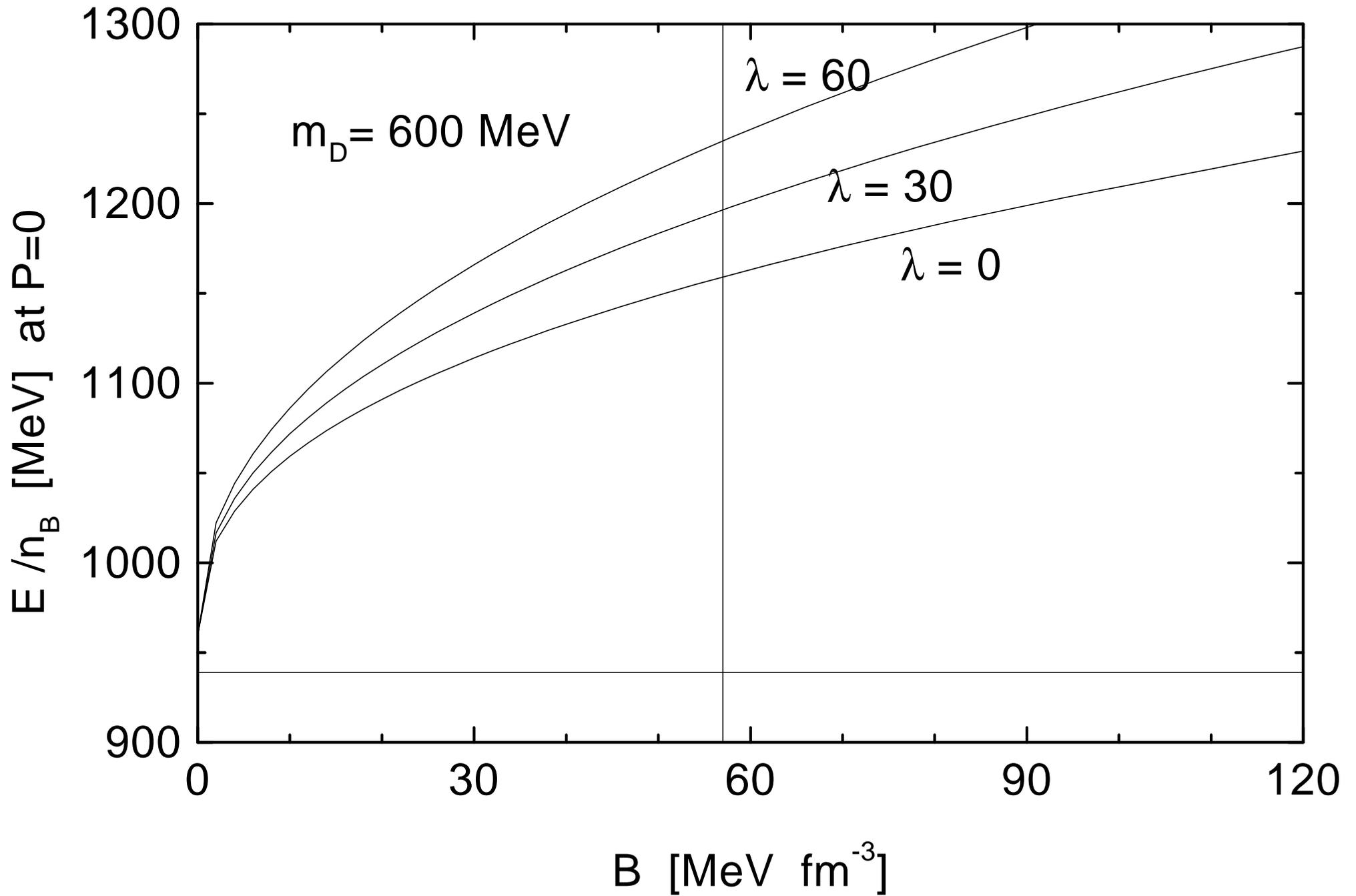
\end{document}